\documentclass[12pt]{iopart}

\usepackage{graphicx}
\begin{document}

\title{Interaction of strangelets with ordinary nuclei}

\author{L Paulucci$^{1,2}$ and J E Horvath$^{2}$}
\address{$^{1}$ Instituto de
F\'\i sica - Universidade de S\~ao Paulo\\
Rua do Mat\~ao, Travessa R, 187, 05508-090, Cidade Universit\'aria\\
S\~ao Paulo SP, Brazil}
\address{$^{2}$ Instituto de Astronomia, Geof\'\i sica e
Ci\^encias
Atmosf\'ericas - Universidade de S\~ao Paulo\\
Rua do Mat\~ao, 1226, 05508-900, Cidade Universit\'aria\\
S\~ao Paulo SP, Brazil}
\ead{paulucci@fma.if.usp.br} 

\begin{abstract}
Strangelets (hypothetical stable lumps of strange quark matter) of
astrophysical origin may be ultimately detected in specific cosmic
ray experiments. The initial mass distribution resulting from the
possible astrophysical production sites would be subject to
reprocessing in the interstellar medium and in the earth
atmosphere. In order to get a better understanding of the claims
for the detection of this still hypothetic state of hadronic
matter, we present a study of strangelet-nucleus interactions
including several physical processes of interest (abrasion,
fusion, fission, excitation and de-excitation of the strangelets),
to address the fate of the baryon number along the strangelet
path. It is shown that, although fusion may be important for low
energy strangelets in the interstellar medium (thus increasing the
initial baryon number $A$), in the earth atmosphere the loss of
baryon number should be the dominant process. The consequences of
these findings are briefly addressed.
\end{abstract}

\pacs{21.65.Qr, 96.50.sb, 96.50.sf}
\submitto{\JPG}

\maketitle

\section{Introduction}

The hypothesis of stability for strange quark matter (SQM)
\cite{Wit, Bodmer, Chin, Terazawa}, a cold form of QCD plasma
composed of u, d, and s quarks, naturally leads to the
consideration of its existence in the interior of neutron stars
\cite{Alcock, SS2, SS3, Alpar, SS6}. It has been conjectured that
if this is indeed the case, then some astrophysical events could
eject finite pieces of SQM, called strangelets, in the Galaxy.
Thus, the latter would ``contaminate'' the cosmic ray flux in the
interstellar medium (ISM), forming an exotic component among
ordinary nuclei (however, see \cite{WK, Janka} for definite
counterexamples from strange stars merging simulations).

Although the density of matter in the ISM is very low (on average
about 1 particle/cm$^3$), the confinement times of charged
particles in the galactic magnetic fields are rather large, so
that collisions between cosmic rays and particles composing the
medium may become a relevant factor. Particularly, the passage of
strangelets through regions in which the density of the ISM is
substantially higher, for example, HII regions and supernovae
remnants, could exert a measurable influence on the ultimate
detection of the flux at a fixed mass value.

On the other hand, the density in the terrestrial atmosphere is at
least fifteen orders of magnitude higher than the average one
found in the ISM. Cosmic rays, thus, must travel a thick air layer
before hitting the earth surface. If the reprocessing of the mass
distribution in the ISM seems likely, it is unavoidable in the
atmosphere, at least for strangelets penetrating deeply in the
atmosphere.

There have been attempts of explaining some rare events in terms
of the presence of strangelets in the cosmic ray flux
\cite{ET, Price, Data1, Pol1, Data3, Boiko, Banerjee2000}.
Those events presented one or more exotic
features such as low charge-to-mass ratio,
high penetration of the primary in the atmosphere, absence of
neutral pion production, transverse moment of secondaries much higher
than the typical values of ordinary nuclear fragmentation, and exotic
secondary production.

Broadly speaking, there are two proposed explanations for the processing
of strangelets (assumed to be primaries, see \cite{Angelis})
that travel deep in the atmosphere,
particularly those arriving with ultrarrelativistic energies.

It has been suggested that strangelets of high baryon number {\it
lose mass} in successive interactions with ordinary nuclei from
the top of the atmosphere during their propagation towards the
earth's surface. When their baryon number reaches a value for
which their mass is lower than the minimum one associated with
stability, $A_{crit}$, they decay into ordinary hadrons. In this
way, it would be possible to conciliate the small mean free path
(of same order of ordinary nuclei) and the high penetration in the
atmosphere \cite{menos}.

On the other hand, in contrast to ordinary nuclei which tend to
fragment in collisions, it has been suggested that strangelets
could become more bound through the {\it absorption} of matter.
Also, the electric charge of strangelets allows their trajectories
to be affected by the geomagnetic field, causing an increase in
the true length of the path taken to reach a given altitude. If
this is the case, then the number of interactions strangelets
would suffer with atmospheric nuclei when reaching a given
altitude would increase, probably resulting in its complete
evaporation before reaching the desired altitude. Therefore, it
has been proposed that strangelets with baryon number slightly
above $A_{crit}$ when reaching the top of the atmosphere would
increase its mass instead of decreasing it, due to successive
fusion reactions with atoms in the atmosphere \cite{mais}. The
possibility of competing fission was not considered in these
models.

Since there are still controversies about these phenomena
by different authors, a more complete analysis is
necessary to provide a definitive answer on what process of
interaction could lead to an event with characteristics similar
to the Centauro and a few other abnormal events.

In this paper we will discuss by which means the initial
fragmentation spectra for strangelets coming from astrophysical
sources could be reprocessed through collisions with the matter
that compose the ISM. We will also discuss how the same hadronic
interactions to which strangelets would be subject in the ISM can
be also responsible for experimental signatures, possibly
detectable after their passage through the earth's atmosphere.

\section{Hadronic Interactions}\label{IntHad}

For the analysis of nuclear interactions between strangelets and
other nuclei, we considered the following processes possibly
affecting the initial baryon number of the strangelet: fusion,
abrasion and fission. We also considered the processes of energy
loss (de-excitation), which relieves the excitation energy
acquired in the collision. Each one is described as follows:

\subsection{Abrasion}

The {\it abrasion} model proposed by Wilson {\it et al}
\cite{Wilson} is a simple model built to describe spallation
qualitatively. It is based on geometric arguments rather than on
details of particle-particle interaction. The abrasion process
consists in the sheer-off of the region of overlap in the target
by the projectile, possibly leading to the removal of all matter
affected by the collision.

In spite of the evident loss of details, this model is very useful
when there is lack of experimental data or when other models fail
to give a consistent analysis. The use of this simplified model
largely based on geometric arguments in this study becomes a
natural choice since strangelets have not been detected yet and,
consequently, models cannot be improved by comparison with
experimental data.

The quantity of matter in the projectile that is abraded is
given by

\begin{equation}\label{abr}
\Delta_{abr}=F\,A_p[1-\exp(-C_T / \lambda)],
\end{equation}

\noindent{where $F$ is the fraction of the projectile in the interaction zone,
$\lambda$ is the mean free path of nucleon-nucleon interaction,
$A_p$ is the baryonic mass of the projectile and $C_T$ is the
maximum chord in the projectile-target
interface (for details, see \cite{Wilson} and references therein).}

To study the interaction of strangelets with ordinary nuclear
matter, we considered parametrically the formation of
``pseudo-baryons'', or clusters of three quarks temporarily bound
in the interior of a strangelet, in a similar procedure to that
used in the literature for the study of the difference in the
binding energy between the nuclei of $^3$He and $^3$H (see, for
instance \cite{Koch}). It is assumed that quarks in the strangelet
will maintain their identities as long as the distance between
them is higher than a certain length scale, $r_0$. Whenever there
is a superposition of the quarks ($r < r_0$),  they are treated as
confined to the same spherical region. We did not consider the
formation probability of clusters other than those made of three
quarks, for we are interested in analysing the change in baryon
number of strangelets. This would not happen in the case of
abrasion of mesons nor with the ejection of a single quark (due
the effect of the colour field). We also did not take into account
the abrasion of clusters containing a higher number of quarks (for
example, a pentaquark) for they would have an extremely small
formation probability, thus being irrelevant for the present
analysis.

We considered that at a given time, a fraction of the quarks
composing the strangelet are grouped in three quarks temporarily
bound. We analyse the results taking this fraction of clusters $f$
as a free parameter. The mean free path is taken according to a
parametrisation used for the nucleon-nucleon cross section
\footnote{In the additive quark model, the cross section for a
given particle is proportional to the number of valence quarks
which compose it. It is known that different baryon compositions
affect the value of the cross section. However, this first
approximation provides the correct order of magnitude for the mean
free path.} \cite{Dymarz} and the abrasion model is used to
estimate the change in the baryon number for each collision
process.

\subsection{Fusion}

The fusion process is represented generically by $A + B
\rightarrow C$, where $A$ and $C$ are strangelets and $B$, the
incident nucleus.

For a fusion reaction to occur, the interacting nuclei must have
enough energy to overcome the repulsive Coulombian barrier between
them, or it can also penetrate the barrier through the well-known
quantum tunnelling effect.

For centre of mass energies lower than the Coulombian barrier of
the system, we simply used the Gamow parametrisation. For energies
above the Coulombian barrier, we followed the proposal of
\cite{JorgeSpall} and consider that when the energy deposited by
the projectile into the strangelet (in the reference frame of the
latter) is of order of the binding energy of the projectile in the
strange quark matter, then fusion occurs. We also consider that
fusion will occur only for central collisions, according to the
geometric parameters defined for the abrasion process. In this
way, for the smallest chord moved along by the interacting nucleus
in the strangelet in a central collision (equal to
$2\sqrt{2R_{str}r_p-r_p^2}$, where $R_{str}$ and $r_p$ are the
strangelet's and nucleus' radius, respectively) we considered the
maximum amount of energy deposited to be equivalent to the binding
energy of the strangelet that fuses ($E_{bound}\simeq
M_n-M(A+A_p)/(A+A_p)$, where $M_n$ is the mass of the neutron and
$M(A)$, the mass of the strangelet of baryon number $A$). A
scaling is taken for higher lengths (up to the strangelet's
diameter). This construction allows to associate each chord of
interaction with a step function in energy for fusion, and is also
consistent with the overall geometric approach adopted from the
beginning.

\subsection{Fission}

The {\it fission} of strangelets may follow after processes
transferring enough excitation energy in a collision.
As in previous studies \cite{Wu}, we considered, in analogy to
ordinary nuclei, a liquid drop model to address this issue.

When the distance between fragments 1 and 2 coming from the
possible fission of a strangelet is $r=0$ (the initial state of
the spherical drop), $E_0$ is the difference in the rest
energy (quantity of energy available for fission) given by

\begin{equation}\label{fis1}
E_0={M(A,Z)-M(A_1,Z_1)-M(A_2,Z_2)},
\end{equation}
\noindent
where (A,Z) represents the strangelet which may come to fission.

The smallest energy necessary for a system to fission (activation
energy) is the one leading to the fragmentation of the system in
two fragments of equal mass. This is easier for strangelets than
for heavy nuclei because the Coulomb energy does not play an
important role. In fact, we neglected the contribution of Coulomb
energy in our calculations, since it is much smaller than the
masses of strangelets themselves.

\subsection{Excitation of strangelets}

Even peripheral collisions not stripping baryon number from the
strangelet can be significant for the excitation of the latter.
The mean energy transferred to a nucleon by unit of intersected
path is of order 13 MeV/fm. As strangelets are composed by quarks,
it is reasonable to assume that the interaction of these particles
with ordinary nuclear matter behaves similarly to that between two
nuclei. In this way, the energy deposited per unit path must be of
the same order of magnitude of the one between nucleons. In this
work, the excitation energy due to the transfer of kinetic energy
through the surface of the interacting system was taken as the
nuclear value and scaled by the longest chord in the surface of
projectile (for more details, see \cite{Wilson}).

In cases in which {\bf abrasion} occurs, the excitation energy
coming from the distortion of the nucleus must also be considered
(the parametrisation used is detailed in \cite{Wilson}).

The SQM hypothesis suggests that the {\bf fusion} of a proton
with SQM leads to the additional liberation of energy for each
nucleon absorbed. In the fusion process the excitation energy
can be computed using the ``mass excess'', $E_x$, written as

\begin{equation}\label{exfus}
E_x=M_N(A_N)+M_{str}(A_{str})-M_{str}(A_{str}+A_N),
\end{equation}
\noindent
being $A_{str}$ and $A_N$ the baryon numbers of strangelet and the
nucleus with which it interacts and $M_N(A)$ and $M_{str}(A)$, the
masses of ordinary nuclear matter and strange matter for a given
$A$, respectively.

\subsection{De-excitation}\label{Desex}

A strangelet may suffer de-excitation through surface evaporation
of nucleons. For the emission of neutrons, we followed the
procedure detailed by Berger and Jaffe \cite{Berger}. The emission
of neutrons through the weak force leaves a strangelet with
parameters changed by $\Delta A=-1$ and $\Delta Y=\Delta Z=0$.
This happens when

\begin{equation}
\frac{\partial E}{\partial A}> M_n.
\end{equation}

On the other hand, the energy lost by the emission of pions can be calculated
based on the chromoelectric flux tube model \cite{Banerjee83}
and reads

\begin{equation}
\frac{dE}{dt}=-1,12\times 10^{20}\,A^{2/3}T^2\exp{(-381,1/T)} \rm{ MeV s}^{-1}.
\end{equation}

It is known that SQM is a poor emitter of thermal photons
at energies below 20 MeV. When considering bremsstrahlung process,
the emission from the surface of SQM is of four orders of magnitude smaller
then the equilibrium black body emission at a given temperature
$T$ \cite{UsovBB}. The equation to be considered is then

\begin{equation}
C_v\frac{dT}{dt}=-\zeta 4\pi R^2\sigma T^4,
\end{equation}
\noindent
where $\zeta\sim 10^{-4}$, $C_v$ is the specific heat of SQM
and $\sigma$ is the Stefan-Boltzmann constant.

Finally, cooling by neutrino emission produces a luminosity
$L_{\nu} = dE_{\nu}/dt$ given by

\begin{equation}
L_{\nu}=\frac{4}{3}\pi R^3\epsilon_{\nu},
\end{equation}
\noindent
where $\epsilon_{\nu}$ is the neutrino emissivity.
The specific heat is taken from references \cite{Iwamoto, Jorge91}
for SQM without pairing and from \cite{Blaschke} for CFL SQM.

\section{Interactions in the ISM}\label{numerico}

If produced in astrophysical sites, strangelets would
interact with the ISM matter causing a
reprocessing of the mass distribution injected at the sources.

To analyse all the possible processes of hadronic interaction
described above between ordinary nuclei and strangelets and
actual de-excitation channels, we built a computer code tracking
interactions on an individual basis. Starting from a given
strangelet of baryon number $A$ and kinetic energy $E$, the
following steps are taken:

\begin{enumerate}

\item{\it Setup of the collision:}

\subitem Random sampling of the impact parameter, $b\leq R_N+R_{str}$,
where $R_N$ and $R_{str}$ are the nuclei and strangelet's radius,
respectively. From this parameter, the distance of closest approach
is calculated;

\subitem Random sampling of the excitation energy for the strangelet
resultant from the transference of energy in the collision
(in half of the events, the target is considered
to be in a excited state and the projectile in the other half);

\item{\it Hadronic interaction:}

\begin{description}

\item If the distance of closest approach is higher than the sum of
the radius of the two particles, Coulombian scattering is assumed to
happen;

\item On the other hand, for energies below the Coulomb barrier,
there is a probability of quantum tunnelling, sampled numerically.
For energies above barrier, the criteria for energy deposition are
checked. If conditions are met, fusion is assumed to occur and the
corresponding excitation energy is obtained;

\item If the conditions for fusion are not fulfilled,
then abrasion is considered
according to equation \ref{abr} and the corresponding excitation
energy is obtained. If there are no baryons extracted in
the interaction, scattering is said to have taken place.

\end{description}

\item{\it Fission:}

Fission can only happen if the total excitation energy is above
the activation energy. If this is the case, then the most likely
channel for fission is considered, i. e., the strangelet will
fragment in two daughters with same baryon number and the kinetic
and excitation energies are equally divided between the fragments;

\item {\it De-excitation:} 

The strangelet's temperature
is obtained according to the First Law of
Thermodynamics so that the de-excitation processes can be
evaluated (emission of photons, neutrinos and pions and neutron
evaporation).

\end{enumerate}

\begin{figure}
\begin{center}
\includegraphics[width=0.51\textwidth]{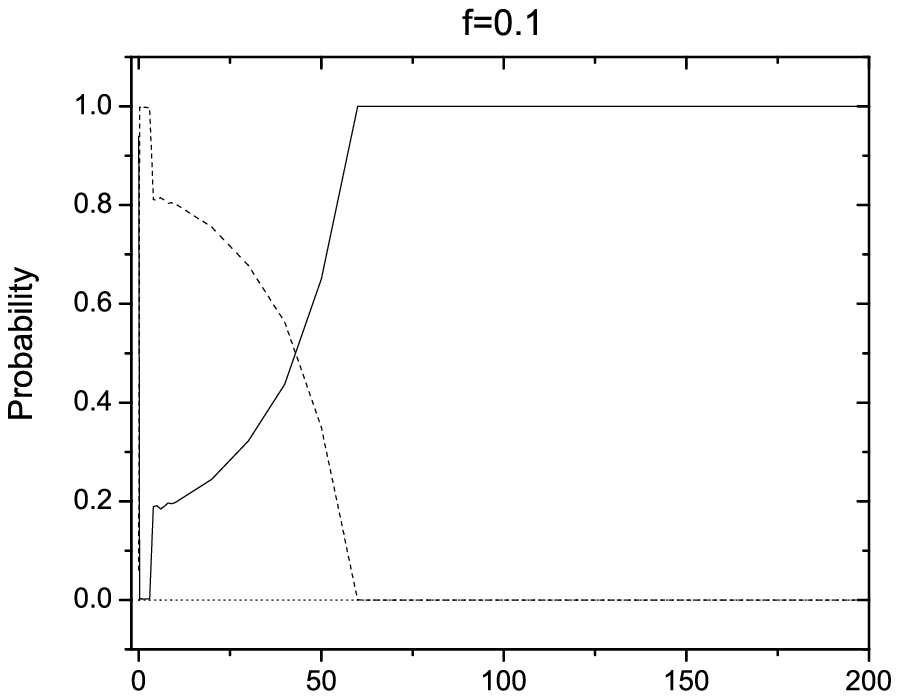}
\includegraphics[width=0.475\textwidth]{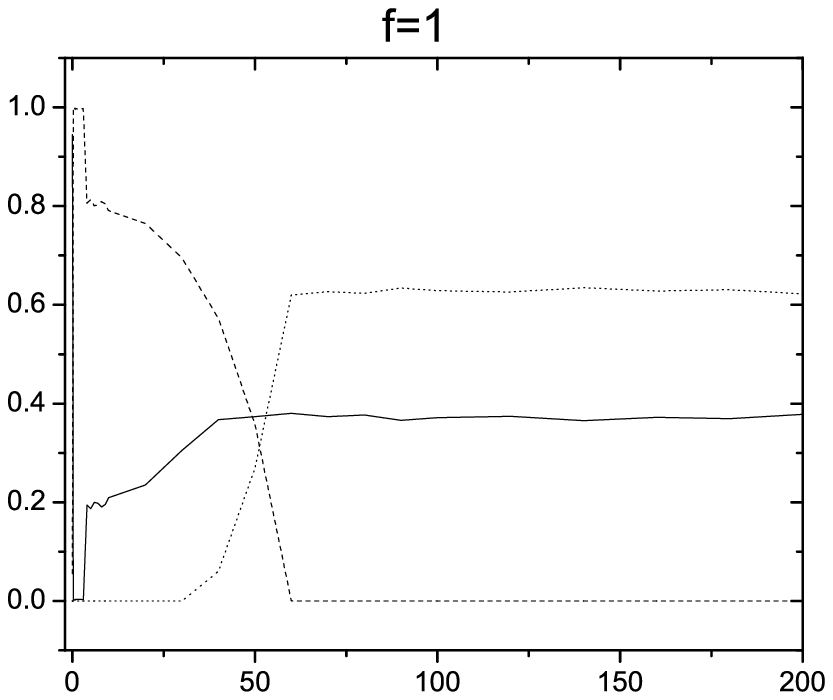}
\includegraphics[width=0.51\textwidth]{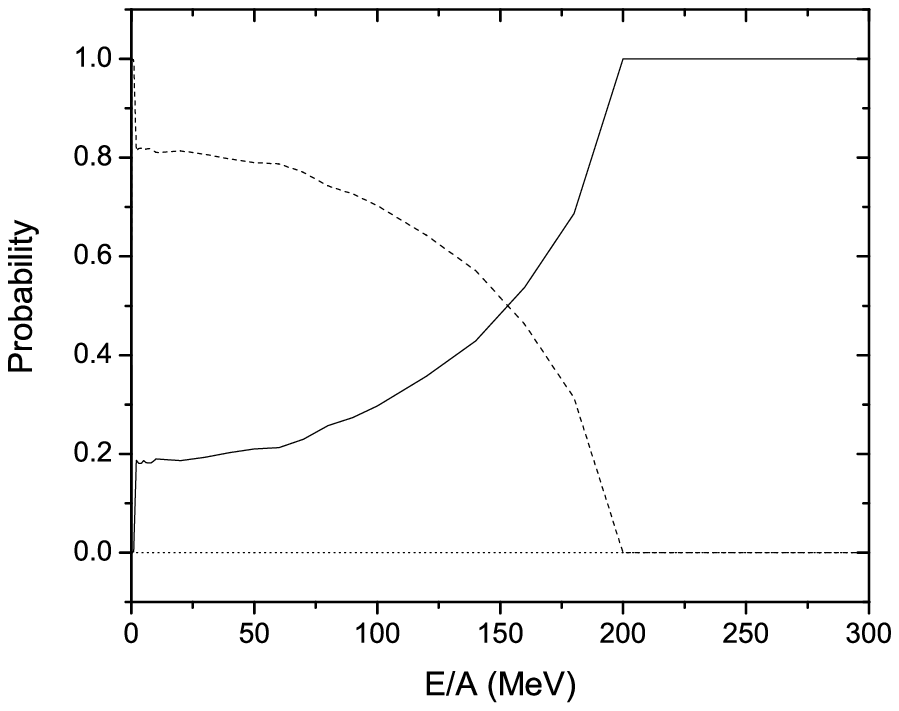}
\includegraphics[width=0.47\textwidth]{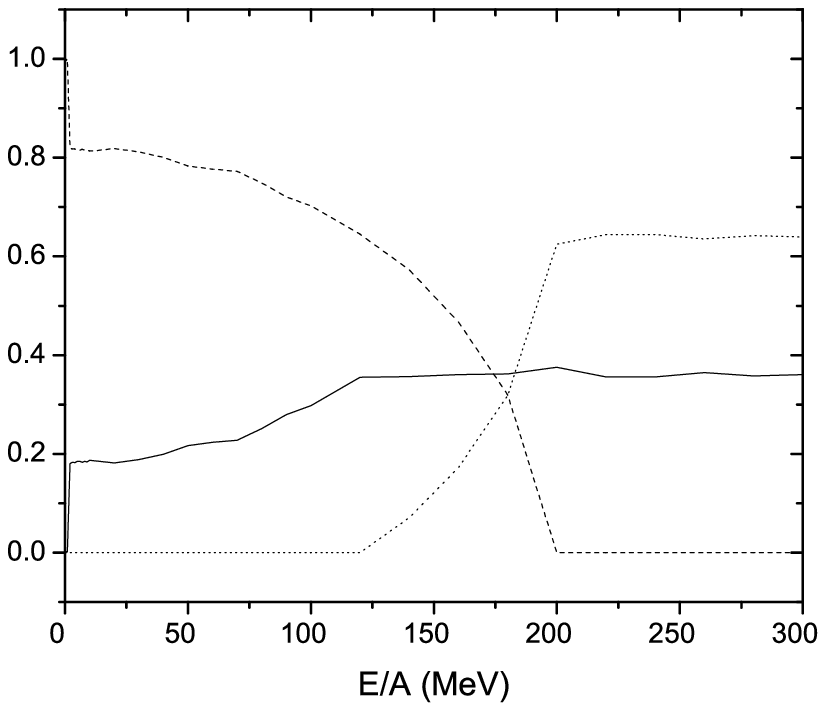}
\caption{Probability of occurrence of hadronic processes for
strangelets with (below) and without (above) pairing of
$A=100$ and fraction of clusters of baryons of 0.1 (to the left)
and 1 (to the right) with protons in the ISM. The full, dashed
and dotted lines are for the processes of scattering, fusion and
abrasion, respectively.}\label{ProcfA100}
\end{center}
\end{figure}

Figure \ref{ProcfA100} presents the probabilities
of abrasion, fusion and scattering
processes, as described in section \ref{IntHad}, for the
collision with protons in the ISM as a function of the
incident energy of the strangelet \footnote{In this analysis
we considered the dependence on the strangelet's mass with the
baryon number according to \cite{strangeletsT}.}.

As discussed above, the fraction of ``baryons'' inside a
strangelet is taken as a free parameter in the calculations and is
related to the probability (impossible to calculate reliably) that
at a given moment there is a certain quantity of clusters of three
quarks formed internally. This value is relevant for verifying the
relative importance of the processes of abrasion and scattering.

For energies below the Coulomb barrier, fusion occurs due to
quantum tunnelling and for energies above barrier, it is the
ultimate result of energy deposition of the projectile in a
central collision. As the energy increases, the probability of a
central collision in which the projectile deposits all its kinetic
energy (in the strangelet's referential) progressively decreases
until the kinetic energy is above the maximum associated to fusion
(the one correspondent to the deposition along the strangelet's
diameter), where the probability goes to zero.

Due to the decrease of the fusion probability as the energy
increases, the abrasion process becomes important. Also the mean
free path decreases with the increase of energy, causing the
collision between the proton and the pseudo-baryons inside the
strangelet to be more probable \footnote{At relativistic energies,
it mimics full stopping of the projectile in the target.}.
Nevertheless, if the fraction of clusters is too low, the abrasion
process becomes dominant only at high energies, since the passage
of the proton through the strangelet does not change its baryon
number.

Although the dependence of the electric charge is less important
for CFL strangelets (which would in turn influence the
determination of the distance of closest approach), for high
enough energies this effect is indeed irrelevant, and this
explains why there are no significant differences between the two
states (unpaired and CFL) in Fig. \ref{ProcfA100}. This fact is
mainly due to the assumption about the mean free path for the
ordinary nucleus with the pseudo baryons formed inside the
strangelet, considered to be the same for both states in this
study. As the number of clusters increases, the influence of the
abrasion process becomes more important with a raise in the number
of baryons abraded. For energies per baryon number high enough
(above $\approx 10^5$ MeV), the relation between abrasion and
scattering processes tends to an asymptotic value.

\begin{figure}
\begin{center}
\includegraphics[width=0.6\textwidth]{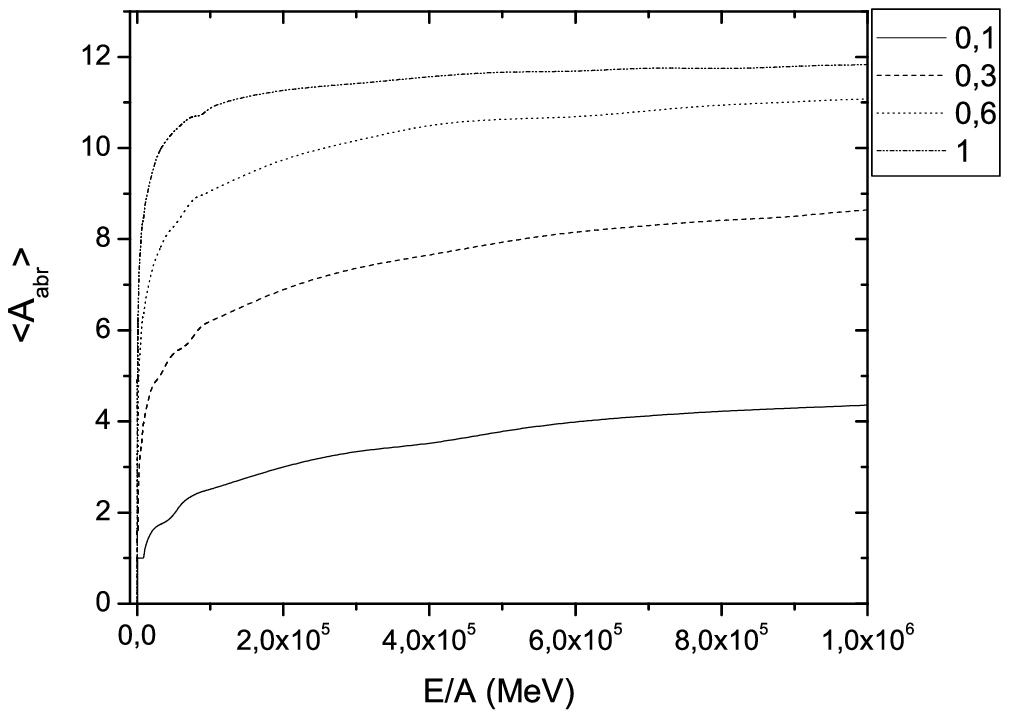}
\caption{Mean abraded matter as a function of energy for
strangelets without pairing in the ISM for $A=1000$ and 
fraction of baryon clusters $f$ as indicated.}\label{Aabr}
\end{center}
\end{figure}

Figure \ref{Aabr} shows the mean baryon number abraded
when abrasion becomes relevant. There are no important differences
in this parameter for non-CFL and CFL strangelets.
As the strangelet energy
increases, the quantity of abraded matter increases until it
approaches an asymptotic value for very high energies.
Obviously, this value is also affected by the fraction $f$ of
clusters temporarily bound in the strangelet. The higher this
fraction, the higher the loss of strangelet mass per collision.

In a collision of a strangelet with ordinary nuclei the excitation
energy must be higher than the activation energy in order to
fission to happen. As mentioned above, the most likely channel for
the fission of strangelets (apart of possible shell effects, not
taken into account here) is the one with two fragments of {\it
same} baryon number (note the difference with ordinary nuclei
influenced by the Coulomb terms). For strangelets interacting with
protons in the ISM the highest energies acquired in the
interaction in this analysis were not enough (by a factor of at
least 10) to cause fission of strangelets of relatively low mass.
Moreover, for high-mass strangelets, fission is again disfavoured
because, although there is a small difference in binding energy
between the strangelet and its fragments, the high baryon number
pumps the activation energies to even higher values than the ones
for low-mass strangelets, as expected.

\begin{figure}
\begin{center}
\includegraphics[width=0.49\textwidth]{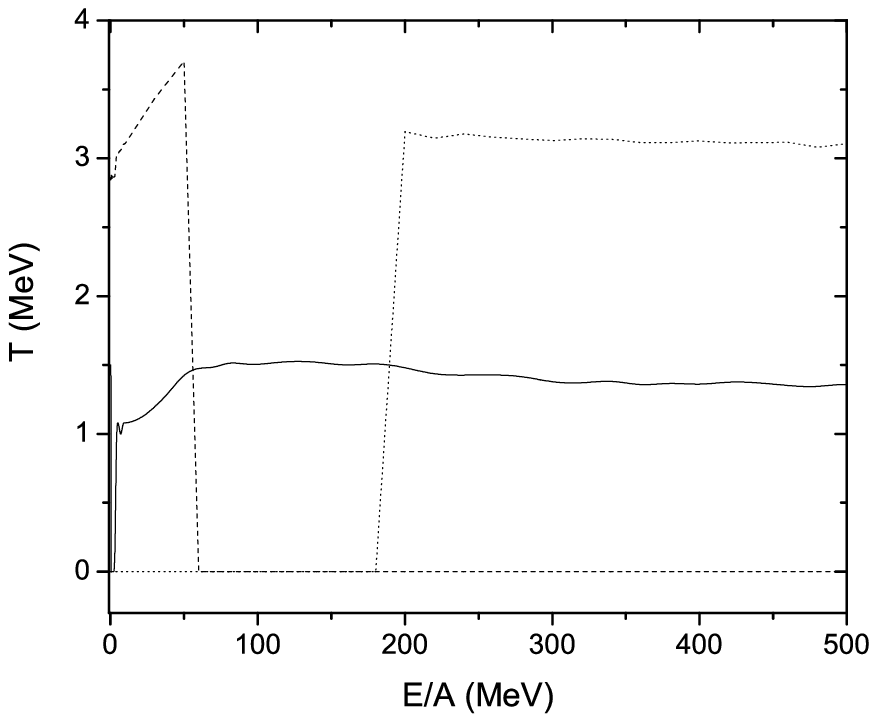}
\includegraphics[width=0.49\textwidth]{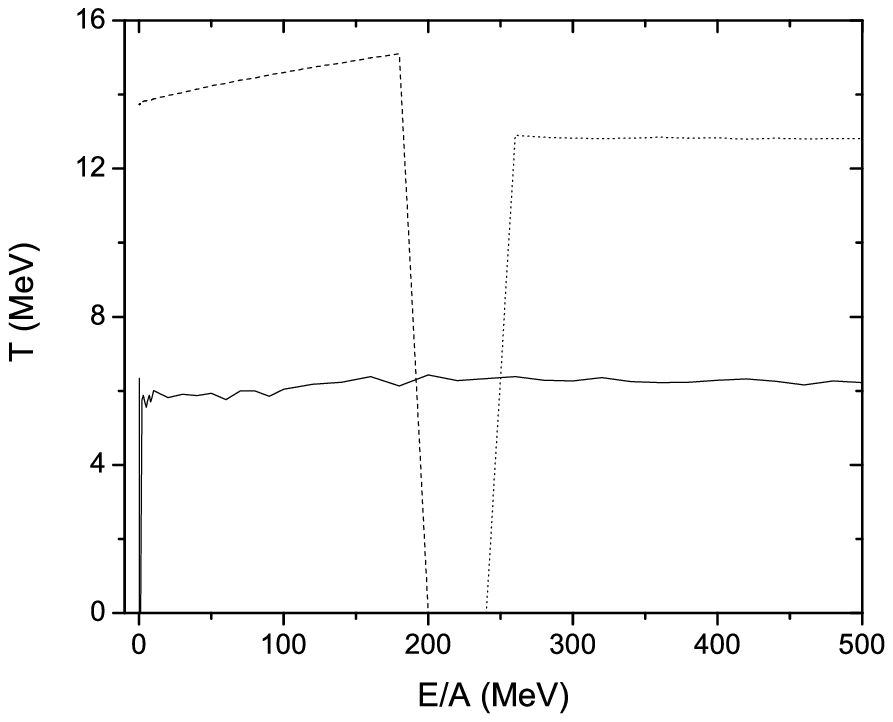}
\caption{Temperature after hadronic interaction of strangelets in
the ISM with (right) and without (left) pairing for $A=100$ and
fraction $f$ of baryon clusters of 0.3, being the full, dashed and
dotted lines for the processes of scattering, fusion and abrasion,
respectively.}\label{Temp}
\end{center}
\end{figure}

We can estimate which processes dominate the de-excitation of
strangelets from their mean temperatures in the collisions (shown
in Figure \ref{Temp}). From the latter, we conclude that the
process of neutron evaporation due to the raise in the temperature
is not likely to be relevant unless the temperatures do rise above
tens of MeV. The emission of neutrons might still be possible
during the pre-equilibrium configuration, while the energy
released during fusion is not still uniformly distributed inside
the strangelet. Nevertheless, and even if this is the case, this
process would probably contribute with the emission of very few
baryons since thermal equilibrium must be reached very quickly.

In addition, cooling by neutrino emission is hardly the main
mechanism for energy loss since the temperatures associated with
the collisions of protons and strangelets are always below a few
MeV, i. e., before neutrino emission dominates the photon
emission. Those temperatures are obviously not enough for pion
emission either. Therefore, we conclude that the dominant
de-excitation mode must be photon emission.

\section{Interactions in the earth atmosphere}

The interactions of strangelets with the main atmospheric
component, the nitrogen molecule, has been analysed afterwards.
We have not considered the possibility of partial fusion.

The same procedure of section \ref{numerico} was performed to
evaluate the relative importance of the abrasion, fusion, fission
and scattering processes of strangelets travelling in the earth's
atmosphere.

\begin{figure}
\begin{center}
\includegraphics[width=0.51\textwidth]{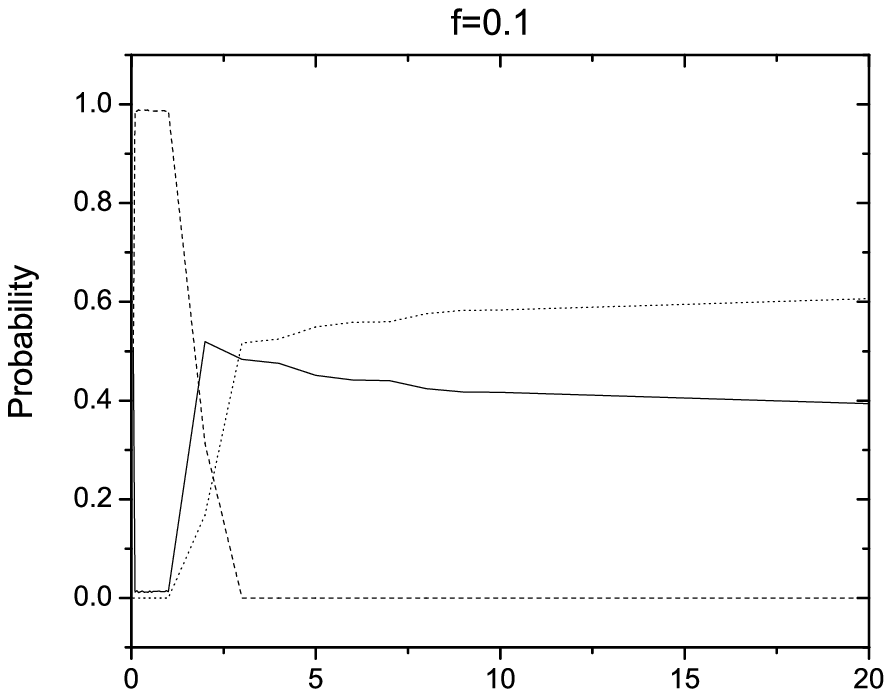}
\includegraphics[width=0.475\textwidth]{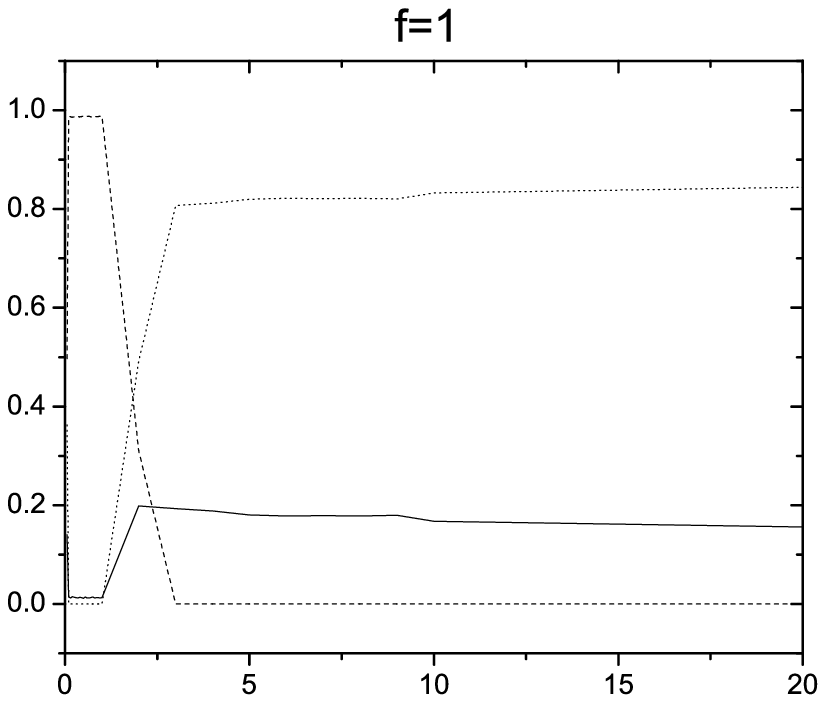}
\includegraphics[width=0.51\textwidth]{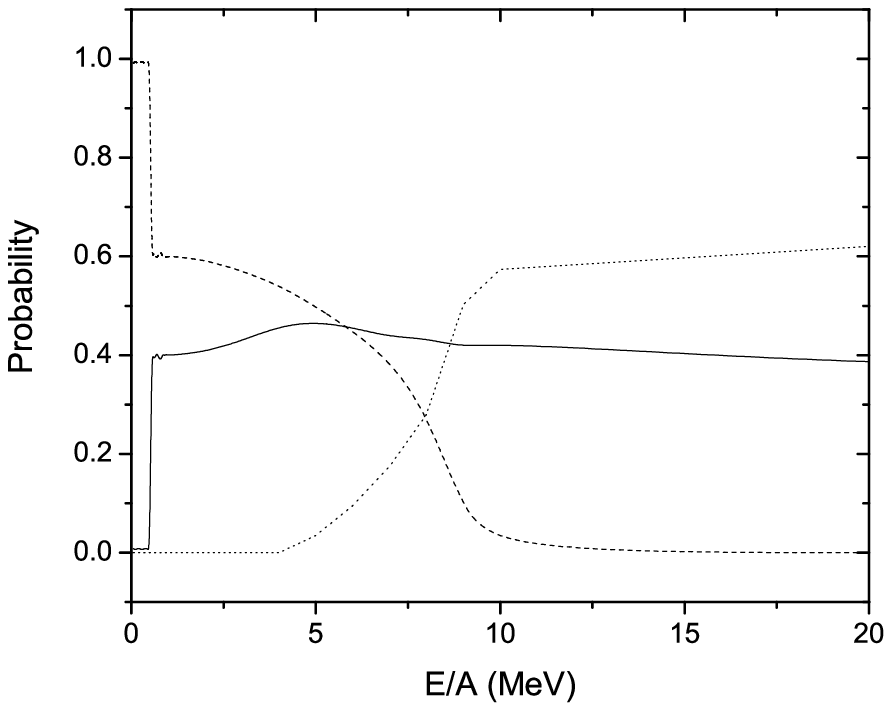}
\includegraphics[width=0.465\textwidth]{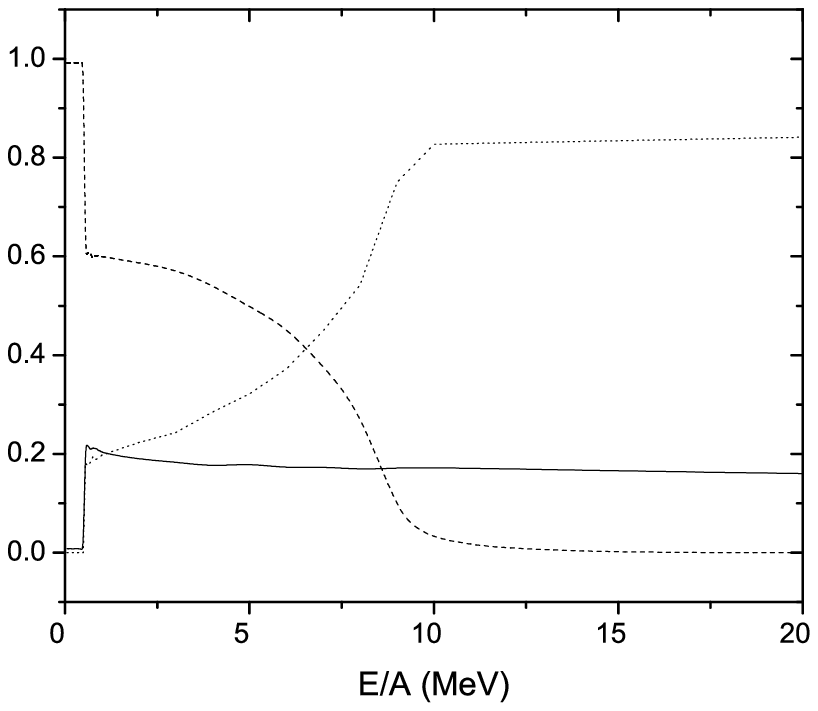}
\caption{Probability of occurrence of hadronic processes for
strangelets with (below) and without (above) pairing of
$A=100$ and fraction of clusters of baryons of 0.1 (to the left)
and 1 (to the right) with atmospheric nitrogen. The full, dashed
and dotted lines are for the processes of scattering, fusion and
abrasion, respectively.}\label{ProcfA100Atm}
\end{center}
\end{figure}

\begin{figure}
\begin{center}
\includegraphics[width=0.5\textwidth]{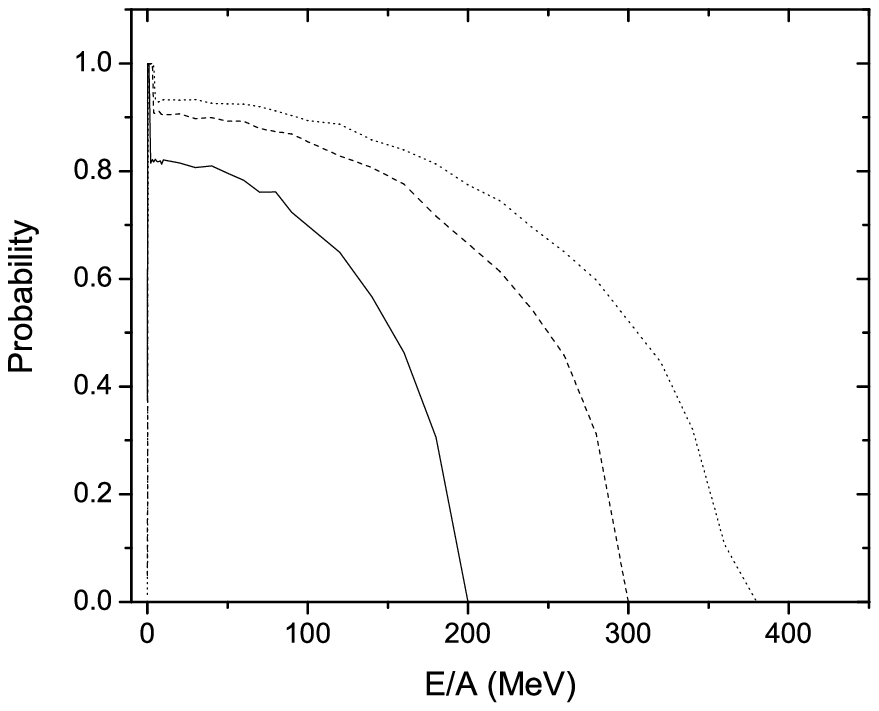}
\includegraphics[width=0.47\textwidth]{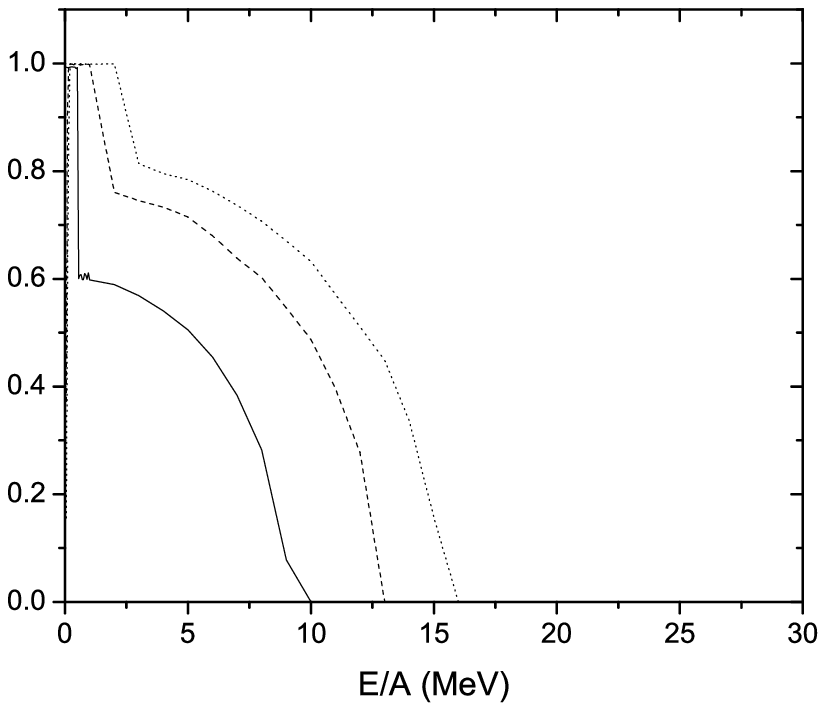}
\caption{Comparison of the fusion probability for CFL strangelets
in the ISM (left panel) and in the atmosphere (right panel)
as a function of incident energy with fraction of clusters of baryons of 0.3.
The full, dashed and dotted lines are for $A=100$, $A=1000$ and
$A=3000$, respectively.}\label{nA}
\end{center}
\end{figure}

\begin{figure}
\begin{center}
\includegraphics[width=0.52\textwidth]{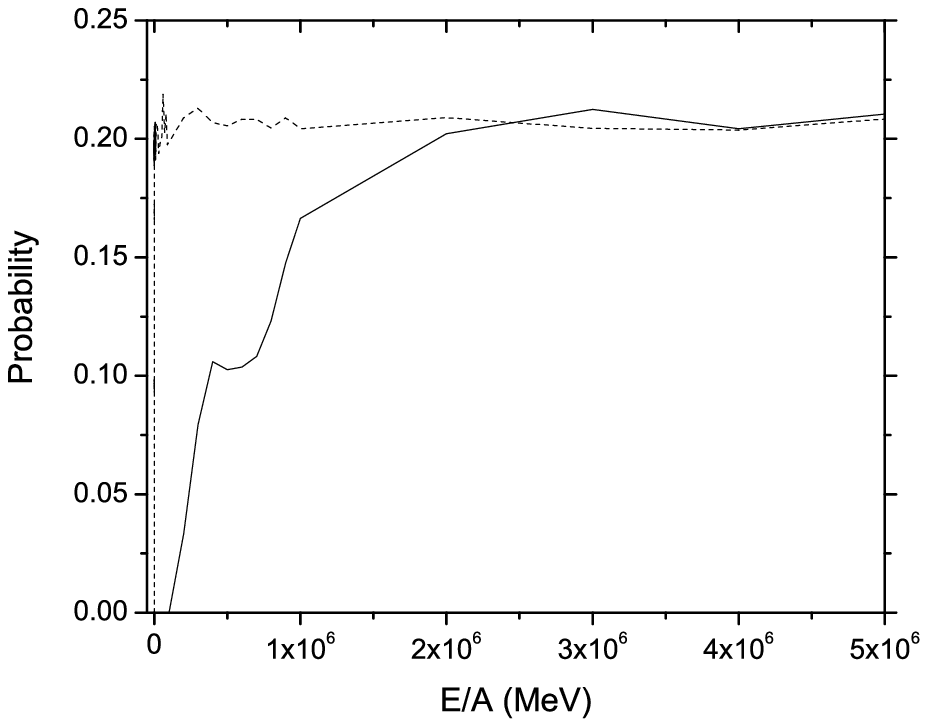}
\includegraphics[width=0.47\textwidth]{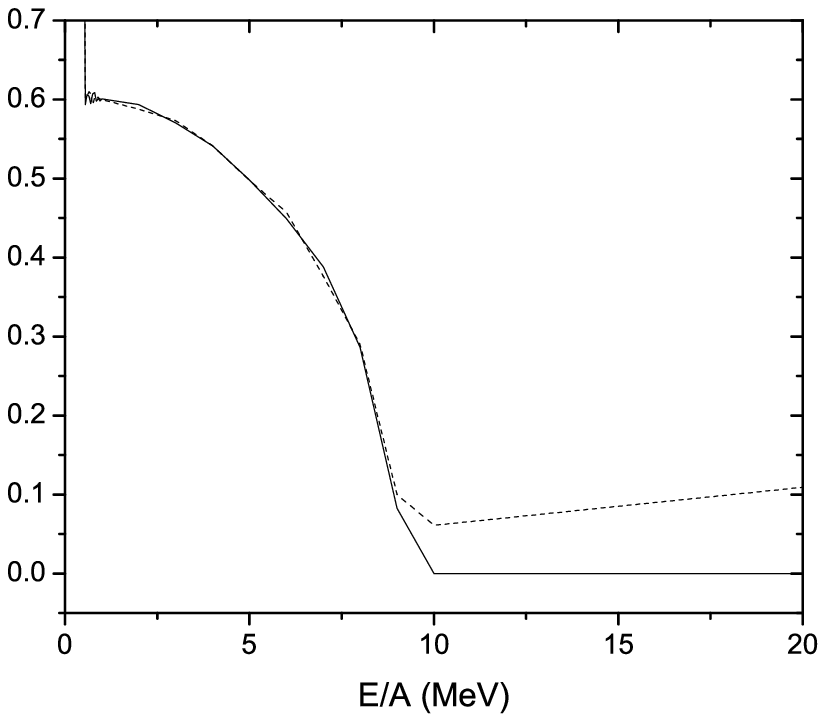}
\caption{Probability of fission as a function of incident
energy for strangelets of baryon number $A=100$ with (to the right)
and without (to the left) pairing for interaction with atmospheric
nitrogen. The fraction of clusters of baryons of 0.1 and 1 are
represented by the full and dashed lines, respectively.}\label{FisAtm}
\end{center}
\end{figure}

The analysis shows that the fusion process only happens for
strangelet energies lower than those in the ISM, due to the
substantial difference in the rest masses of the proton and
nitrogen (see figure \ref{nA}). The fusion probability increases
with $A$ because the increase of the strangelet radius leads to a
longer path in central collisions for the deposition of kinetic
energy of the nucleus with which the strangelet interacts (in the
reference frame of the latter).

The probability for abrasion remains low for low energies,
due to the competition with the fusion process, and increases
with the raise in energy because, with the reduced influence of
the Coulombian sheer off, scattering becomes less likely to happen.
These observations are weakly dependent on $A$.

The abrasion process probability increases with energy and the
quotient abrasion/scattering tends asymptotically to a constant value (close to
1) for relativistic strangelets due to the dependence of the
mean free path of nucleon-nucleon interaction with energy.
This observation is independent of both $A$ and the different
fractions of clusters temporarily formed inside the strangelet.

The mean excitation energy for each process after collisions
is up to $\sim 800$ MeV for abrasion (depending weakly
on $A$ and independent of the state of SQM within the considerations
adopted in this work) and up to $\sim 1.3$ GeV and $\sim 3$ GeV
for strangelets without pairing and CFL, respectively, after
fusion of the strangelet with nitrogen. These values justify the
observation of induced fission for CFL strangelets of low mass
(see figure \ref{FisAtm}). For low energies, fission occurs
due to the liberation of energy from fusion (both probability curves
can be superimposed, see figures \ref{nA} and \ref{FisAtm}).

For higher energies, fission of strangelets with and without
pairing happens due to the excitation energy available from the
deformation of the strangelet as the result of abrasion of a quite
high baryon number $A^{*}$. Although the mean excitation energy in
the abrasion process is smaller than the activation energy, a
fraction of those events can reach energies for which fission is
allowed. In these cases, the probability is smaller for lower
fractions of pseudo-baryons since it would result in a small
quantity of abraded matter (i. e., less distortion of the
strangelet). Nevertheless, this result must be taken with caution
because the abrasion model, established for the description of
ordinary nuclear matter, certainly did not predict the abrasion of
such high amounts of baryons $A^{*}$. For high-mass strangelets,
the total decrease in baryon number can reach values of order
$10^2$, which in turn might lead to overestimated excitation
energies.

The possibility of fission for high-mass strangelets is not
reached for the excitation energies in the collision of these
particles with nitrogen are never sufficient to overcome their
activation energies.

\begin{figure}
\begin{center}
\includegraphics[width=0.52\textwidth]{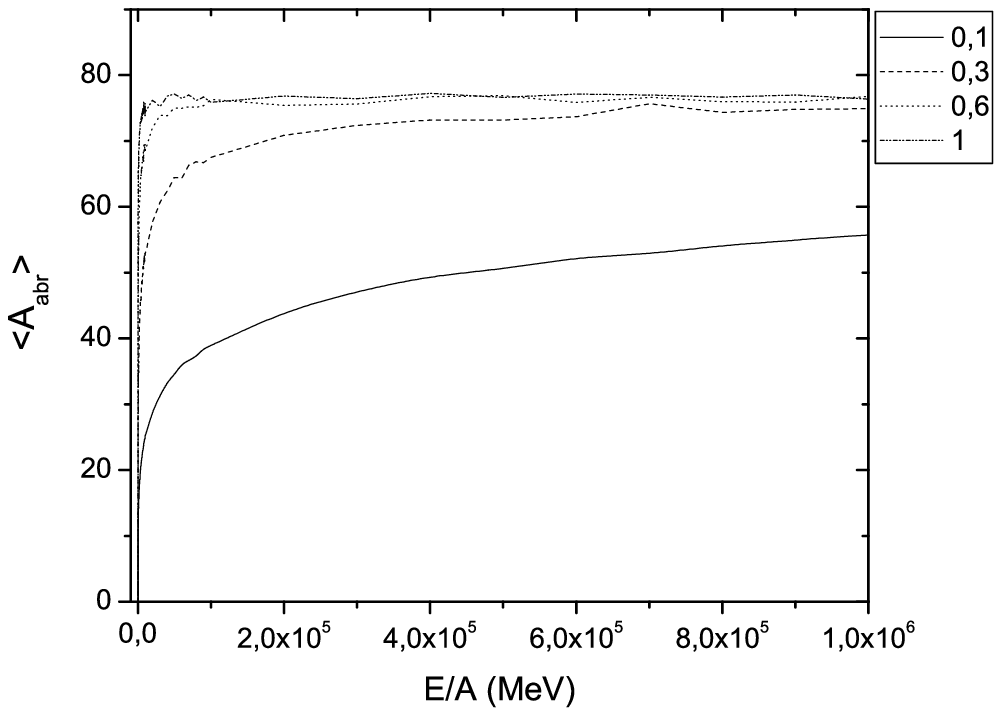}
\includegraphics[width=0.47\textwidth]{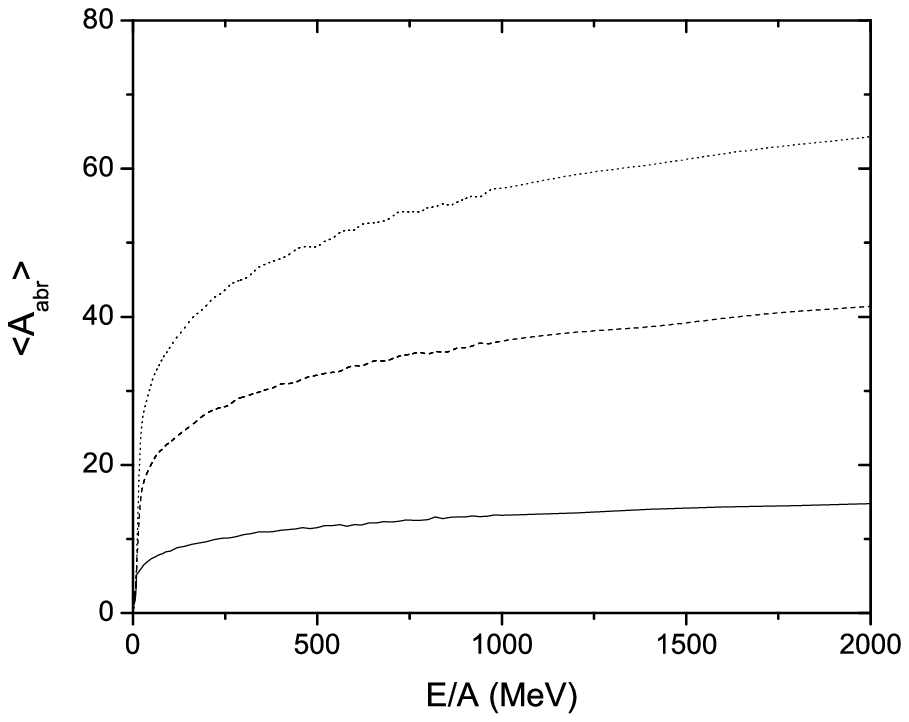}
\caption{Mean abraded matter as a function of energy for
CFL strangelets in the atmosphere. On the left, 
for $A=1000$ and fraction of baryon clusters
$f$ as indicated. On the right, amount of abraded matter for
$f=0.3$ with the full, dashed and dotted curves standing for
$A=100$, $A=1000$ and $A=3000$, respectively.}\label{AabrAtm}
\end{center}
\end{figure}

The mean abraded matter is higher in the atmosphere than in
the ISM since the interaction overlap between the target and projectile
is substantially higher (see figure \ref{AabrAtm}). It also increases
with the fraction of clusters of baryons $f$, but shows
significant differences only for $f=0.1$ for high energies
due to the difference in the mean free path. For low energies,
the difference in the amount of abraded material with the change in $f$ is more
pronounced. These conclusions hold for different baryon numbers
of strangelets and is weakly dependent on the pairing state
of SQM (unpaired or CFL).

The temperatures associated with the fusion process in the atmosphere
are higher than those for fusion in the ISM because,
with complete fusion, a large amount of energy
is liberated {\it per nucleon} of the nitrogen. This allows
fission of CFL strangelets of low baryon number. The temperatures
obtained after fusion are lower than $\sim 6$ MeV for unpaired
strangelets and when fusion is followed by fission, up to
$\sim 18$ MeV for CFL strangelets. For strangelets without
pairing and CFL with $A > 1000$, temperatures from the fusion
process (not followed by fission), which are strongly dependent on the
baryonic content, are always lower than $\sim 3$ MeV and $\sim 14$ MeV, respectively.

The temperatures obtained in the present analysis indicate that
the most likely channel for de-excitation for strangelets
interacting in the atmosphere is photon emission. For the highest
possible temperatures, emission of pions is also possible, but
with a less effective contribution than photon de-excitation. Even
at the highest temperatures, neutron emission is not important
after thermal equilibration.

\subsection{Atmospheric showers initiated by strangelets}

Strangelets penetrating deeply in the atmosphere may lead to the
generation of showers. Note that we have not considered the
formation of ``SQM'' (possibly metastable due to the high
temperatures) in collisions of ordinary cosmic rays with
components of the atmosphere \cite{Angelis}.

To evaluate the possible development of the interaction
of strangelets penetrating the atmosphere from the results
shown in the previous section, it is also necessary to address the
geomagnetic field influence on those particles.

Strangelets with kinetic energies per baryon number lower than
hundreds of MeV to some GeV will have their flux at the surface of
the earth lowered due to the local geomagnetic cutoff. Also, if
the rigidity is above the local cutoff but $E/A$ is of order of
tens to few hundreds of MeV, depending on the strangelet mass, it
is more likely that they will become trapped in the geomagnetic
field lines, if their pitch angles are adequate
\cite{Paulucci2008}. Therefore, the possibility of showers
generated by low energy strangelets is substantially reduced.
Also, the possibility that the fusion process is the one allowing
penetration of strangelets to low altitudes seems very unlikely
because the results of the previous section indicate that fusion
should be important precisely for these low energy strangelets
affected by the magnetic field.

\begin{figure}
\begin{center}
\includegraphics[width=0.49\textwidth]{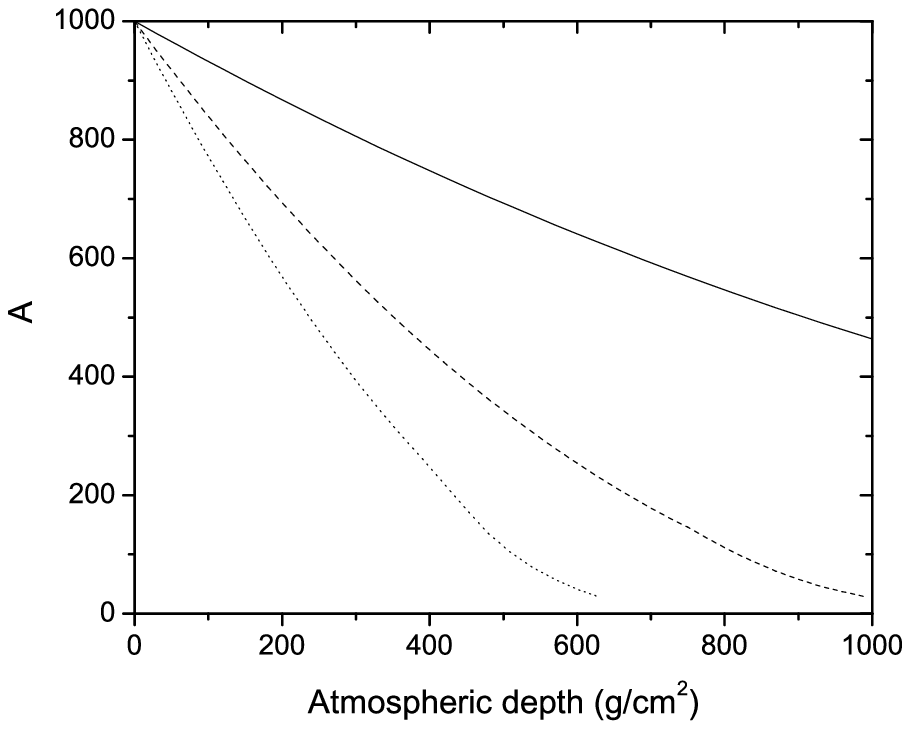}
\includegraphics[width=0.49\textwidth]{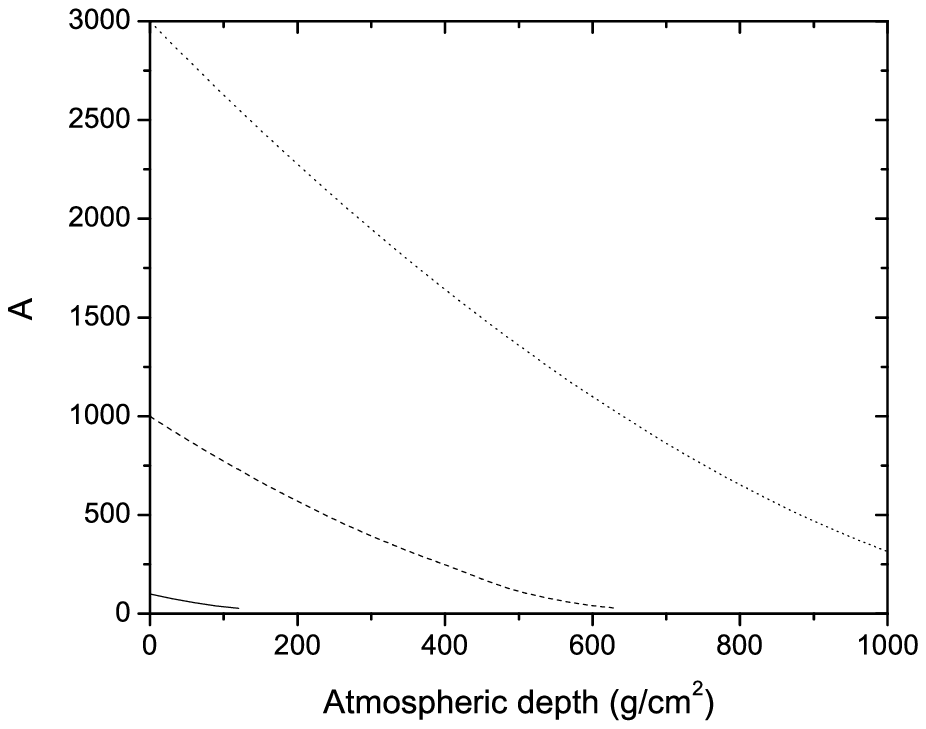}
\caption{Baryon number evolution for a strangelet as a function
of the column density traversed in the earth atmosphere. These values
are not significantly changed with the pairing state of strangelets
in this work. On the right, the strangelet's
baryon number is fixed to $A = 1000$ and the full, dashed and dotted curves
are for $E/A = 100, 10^4, 10^6$ MeV, respectively. On the left, the
strangelet's energy per baryon number is fixed in $E/A = 1 \times 10^6$
MeV and the full, dashed and dotted lines stand for $A = 100, 1000, 3000$,
respectively.}\label{AtmPenet}
\end{center}
\end{figure}

If we assume that the column density travelled between collisions
of strangelets with nuclei in the atmosphere is of order $30$
g/cm$^2$, a value appropriate when $R_{strang}\sim R_{ar}$
\footnote{For strangelets with high baryon number, the values
obtained for the final mass after a given column density traversed
must be overestimated then, since their radius are big when
compared to the nitrogen radius. In this sense, the mean distance
between collisions in the atmosphere should be lower than the one
equivalent to $30$ g/cm$^2$ of column density.}, and imposing that
in a typical collision the energy loss is $\Delta E/E\sim 3$\%
\cite{Data1}, we obtained a crude evaluation of the evolution in
baryon number for strangelets as a function of the column density
traversed, shown in Figure \ref{AtmPenet}.

As expected, the steepness of the mass loss increases for higher
baryon number and/or higher energies at fixed $A$.

It is worth to remark that fission is important for low mass strangelets.
Although less likely than abrasion, this process contributes to prevent these
particles from penetrating deeply the atmosphere before reaching
the minimum $A$ for which SQM should be stable.

The analysis presented in \cite{Wu} models the possibility for
fission following the process of fusion with air nuclei
of strangelets in the atmosphere taking into account the
contribution of rotational energy. Particles with high values of
deformation by rotation have higher probability of fission in a
collision than spherical particles (without any deformation).
Nevertheless, their spallation estimate is not modelled in detail,
rather considering that the strangelet will lose the same mass
number as the air nucleus in each collision. Besides that, our
results are not too different in absolute values (although the
curves of mass evolution for strangelets do have opposite
curvatures), which leads us to believe that the curves shown in
Figure \ref{AtmPenet} should be actually considered as upper
limits.

When considering the celebrated Centauro events, our results
indicate that the events might be explained by a high-mass and
high energy strangelet penetrating the upper atmosphere and
suffering successive baryon number losses until it reaches ground
experiments. Nevertheless, further analysis related to the
kinematics and multiplicities of the secondary production are
necessary for a firmer conclusion.

\section{Conclusions}

The most important feature of this work is to propose a 
consistent approach which accounts simultaneously for 
different processes of interaction between strangelets and 
ordinary nuclei. The use of the abrasion method allows the 
evaluation of the abraded matter for strangelets, rather 
than assuming a fixed baryon number extracted by the interacting 
nuclei or any other oversimplified assumption (as in, for 
example, \cite{menos, mais, Wu, Madsen05}).

We have shown that the reprocessing in mass number of strangelets
in the ISM is a process that must be effectively operating for
long periods due to the high mean free path for interactions
caused by the low-density of nuclei in the Galaxy.

For the analysis of spallation we used the abrasion model,
strongly rooted on geometrical arguments. In spite of the
generality of the assumptions, this has the advantage of making
the results quite independent of experimental data, which are
non-existent in this present case. This adaptation of the existing
models obviously does not describe the details found in
experiments of nuclear matter collision, but is qualitatively
acceptable, and thus we expect the results presented here to point
towards a general trend for strangelet-nucleus interactions.

We have shown that important differences in the results arise with
the assumptions of the fraction of clustering of quarks between
$f=0.1$ and 1. Obviously, $f=1$ is not realistic at all, since in
this case one could consider the strangelet as a kind of gas of
$\Lambda$ particles, something which is inconsistent with the
strange quark matter stability hypothesis from the very beginning
\cite{Bethe87}. However, and in spite of this uncertainty, the
analysis suggests that adopting spallation with nuclear parameters
as the mechanism for reprocessing of strangelets in the ISM should
overestimate the change in baryon number of the primaries. In addition,
the treatment of spallation presented here can
overcome the problems of considering that this specific process of
interaction simply destroys the strangelet (as
assumed, for example, by Madsen \cite{Madsen05}). Also,
we contend that fusion should be considered as an important
process for interaction with protons at low energies. The
estimates of the reprocessing of the initial mass distribution of
strangelets in the ISM should be reanalysed for the better
prediction of the most likely channels for their ultimate
detection and evaluation of existing upper limits to their flux.
The consistent framework of relevant interactions facing the 
unknown physics of strangelets presented here can provide elements for such
reanalysis.

On the other hand, we believe that in the terrestrial atmosphere,
the dominant mechanism to which strangelets are subject is the
loss of baryon number, mostly due abrasion, but also featuring a
contribution from the fission process.

If strangelets are part of the cosmic ray flux, it would be
possible to detect them, especially in experiments located at the
top of mountains \cite{SLIM} since the mass
loss due to interaction with atmospheric particles tend to be
catastrophic for high column densities crossed. It is not 
ruled out that Centauro events may have their origin in
strangelets, although these suggestive results are still to be
analysed in further details.

\ack{We acknowledge the financial support received from
the Funda\c c\~ao de Amparo \`a Pesquisa do Estado de
S\~ao Paulo. J.E.H. wishes to acknowledge the CNPq
Agency (Brazil) for partial financial support.}

\bibliography{Int}

\end{document}